\documentclass[aps, prd, groupedaddress, preprint, tightenlines,  eqsecnum, showpacs, nofootinbib]{revtex4}

\usepackage{natbib}

\usepackage{epsfig}
\usepackage{graphicx}

\usepackage{float, slashed, graphicx, amssymb, amsmath}
\usepackage[usenames, dvipsnames]{xcolor}

\def\Li{\mathop{\hbox{\rm Li}}\nolimits}

\def\ksl{\slashed{k}}

\def\Fcc{P^{(2)}}

\def\spa#1.#2{\left\langle#1\,#2\right\rangle}
\def\spb#1.#2{\left[#1\,#2\right]}
\def\la{\langle}
\def\ra{\rangle}

\def\CY{Cy}

\DeclareMathOperator{\tr}{\mathrm{Tr}}

\def\be{\begin{equation}}
\def\ee{\end{equation}}

\begin{document}

\hfill\today


\title{The Full Color Two-Loop Six-Gluon All-Plus Helicity Amplitude }

\author{Adam~R.~Dalgleish, David~C.~Dunbar, Warren~B.~Perkins and  Joseph~M.W.~Strong}

\affiliation{
College of Science, \\
Swansea University, \\
Swansea, SA2 8PP, UK\\
\today
}

\begin{abstract}
We present 
the full color two-loop six-point all-plus Yang-Mills amplitude in compact analytic form.
The computation
uses four dimensional unitarity and augmented recursion.
\end{abstract}


\maketitle

\section{Introduction}

Computing perturbative scattering amplitudes in gauge theories is a key tool in confronting 
theories of particle physics with experimental results and there is considerable demand for new predictions particularly 
at ``Next-Next-Leading Order'' (NNLO)~\cite{Bendavid:2018nar,Azzi:2019yne}.  
Amplitudes are also the custodians of the symmetries 
of the theory  and as such are important for exploring 
properties of theories which are not always manifest in a Lagrangian approach. 
Computing amplitudes in closed analytic form is particularly useful in this regard. 

Amplitudes for the scattering of gluons within a gauge theory are key, being both important 
phenomenologically and central to gauge theory.  Modern techniques have driven progress in the calculation of analytic expressions for tree and one-loop gluon scattering amplitudes but analytic expressions for two-loop and beyond amplitudes are relatively rare (although in theories of extended supersymmetry a great deal more progress has been made~\citep{Caron-Huot:2019vjl,Bourjaily:2019gqu}).

Computing two-loop amplitudes for gluon scattering in analytic form has proceeded by separating the amplitude into its physical components. Specifically,
amplitudes with a given color structure and specific choice of external helicities have been computed.   For four-point scattering, 
all of these components have been calculated~\citep{Glover:2001af,Bern:2002tk}  (and more recently to all orders of dimensional regularisation in 
\citep{Ahmed:2019qtg}).  At five-point and beyond,
progress has been made in a variety of stages.  In terms of color structure, the simplest amplitudes are the ``leading in color''  amplitudes which only
require planar two-loop integrals to be computed. For external helicity, the ``all-plus''  amplitude, where all external (outgoing) legs have the same helicity, has the most symmetry and is the simplest. The all-plus amplitudes vanish at tree level and so they have a relatively simple singularity structure at loop level.  
In \citep{Badger:2013gxa,Badger:2015lda}
the five-point all-plus leading in color amplitude was computed using generalised unitarity techniques and subsequently presented in a very simple analytic form~\cite{Gehrmann:2015bfy}.  In ref.~\cite{Dunbar:2016aux}  it was recomputed using simpler four dimensional unitarity and recursion methods which is the methodology we use in this article.  
The remaining leading in color five-point
helicity amplitudes have also been computed: in ref.~\cite{Badger:2018enw} the ``single-minus'' (an amplitude which also vanishes at tree level) was computed and the remaining helicities in \cite{Abreu:2019odu}.
In ref.~\cite{Badger:2019djh}  the remaining parts of the full color all-plus five-point amplitude were calculated.  
Beyond five-point only a few amplitudes are known. The leading in color all-plus amplitudes have been 
computed using our
methodology for six-gluons~\citep{Dunbar:2016gjb} and seven gluons~\citep{Dunbar:2017nfy}.  In ref.~\cite{Dunbar:2020wdh} a conjecture for a specific color sub-amplitude was presented valid for $n$-gluons.

In this article, we compute and present in closed analytic form the full color 
all-plus six-point amplitude ${\cal A}_6^{(2)}(1^+,2^+,3^+,4^+,5^+,6^+)$.  
This is the first full color six-point amplitude and contains a wider
class of color amplitudes than the four- and five-point cases. 
Our methodology involves computing the polylogarithmic and rational parts of the finite remainder by a combination of techniques. 
The polylogarithms are computed using four dimensional unitarity cuts and the rational parts are determined by recursion.  
The amplitude contains double poles in (complex) momenta  and we overcome the concomitant issues by
using augmented recursion~\cite{Alston:2012xd}. Our methods bypass the need to calculate non-planar integrals.

\section{Full Color Amplitudes}

A general two-loop amplitude for the scattering of $n$ gluons in a pure $SU(N_c)$ or $U(N_c)$ gauge theory 
may be expanded in a color trace basis as
\begin{eqnarray}
& & {\cal A}_n^{(2)}(1,2,\cdots ,n) =
N_c^2 \sum_{S_n/\mathcal{P}_{n:1}}  \tr[T^{a_1}T^{a_2}\cdots T^{a_n}] A_{n:1}^{(2)}(a_1,a_2,\cdots ,a_n) \notag \\
&+&
N_c\sum_{r=2}^{[n/2]+1}\sum_{S_n/\mathcal{P}_{n:r} }   \tr[T^{a_1}T^{a_2}\cdots T^{a_{r-1}}]\tr[T^{b_r} \cdots T^{b_n}] 
A_{n:r}^{(2)}(a_1,a_2,\cdots ,a_{r-1} ; b_{r}, \cdots, b_n)  
\notag \\
&+& \sum_{s=1}^{[n/3]} \sum_{t=s}^{[(n-s)/2]}\sum_{S_n/\mathcal{P}_{n:s,t}} 
\tr[T^{a_1}\cdots T^{a_s}]\tr[T^{b_{s+1}} \cdots T^{b_{s+t}}]
\tr[T^{c_{s+t+1}}\cdots T^{c_n}] 
\notag
\\
& & 
\hskip 7.0truecm 
\times A_{n:s,t}^{(2)}(a_1,\cdots ,a_s;b_{s+1} ,\cdots, b_{s+t} ;c_{s+t+1},\cdots, c_n ) 
\notag \\
&+&\sum_{S_n/\mathcal{P}_{n:1}}  \tr[T^{a_1}T^{a_2}\cdots T^{a_n}] A_{n:1B }^{(2)}(a_1,a_2,\cdots ,a_n)\,.
\end{eqnarray}
The partial amplitudes multiplying any trace of color matrices are cyclically symmetric in the indices within the trace.  The
summations count each color structure exactly once. Specifically, 
when the sets are of different lengths ($r-1\neq \frac{n}2$, $s\neq t $, $t\neq\frac{n-s}{2}$ and $3s\neq m,n$)
the sets $\mathcal{P}_{n:\lambda}$ 
are
\begin{align}
\mathcal{P}_{n:1}&=Z_{n}(a_1,\cdots, a_n),\notag\\
\mathcal{P}_{n:r}&=Z_{r-1}(a_1,\cdots,a_{r-1})\times Z_{n+1-r}(a_r,\cdots,a_n),  \;\;\ r > 1, r-1 \neq n+1-r  \notag\\
\mathcal{P}_{n:s,t}&=Z_s(a_1,\cdots,a_s)\times Z_t(a_{s+1},\cdots,a_{s+t})\times Z_{n-s-t}(a_{s+t+1},\cdots,a_n)\,.
\end{align}
When the sets have equal lengths, to avoid double counting
\begin{align}
\mathcal{P}_{2m:m+1}&=Z_{m}(a_1,\cdots,a_m)\times Z_{m}(a_{m+1},\cdots,a_{2m})\times Z_2, \\
\mathcal{P}_{n:s,s}&= Z_s(a_1,\cdots,a_s)\times Z_s(a_{s+1},\cdots,a_{2s})\times Z_{n-2s}(a_{2s+1},\cdots,a_n)\times Z_2,\notag\\
\mathcal{P}_{3m:m,m}&=Z_{m}(a_1,\cdots,a_m)\times Z_{m}(a_{m+1},\cdots,a_{2m})\times Z_{m}(a_{2m+1},\cdots,a_{3m})\times S_{3},\notag \\
\mathcal{P}_{2m:2s,m-s}&= Z_{2s}(a_1,\cdots,a_{2s})\times Z_{m-s}(a_{2s+1},\cdots,a_{s+m})\times 
Z_{m-s}(a_{s+m+1},\cdots,a_{2m})\times Z_2  \,.\notag
\end{align}
For example for $A_{6:2,2}(a,b;c,d;e,f)$ the manifest symmetry is
\begin{align}
\mathcal{P}_{6:2,2}=Z_2(a,b)\times Z_2(c,d)\times Z_2(e,f) \times S_3(\{a,b\},\{c,d\},\{e,f\})
\end{align}
which means the summation of this particular term is over 15 terms.

The above expansion is valid for both a $SU(N_c)$  gauge theory and a $U(N_c)$ gauge theory.  In the expansion for $SU(N_c)$ the color trace terms with a 
single trace $\tr[T^{a_i}]$ are omitted.  Specifically these are the terms $A_{n:2}^{(2)}$ and $A_{n:1,s}^{(2)}$ and $A_{n:1,1}^{(2)}$.
These functions are consistent gauge invariant objects whose role is the cancel other terms.
By letting one or more of the external gluons lie in the $U(1)$ part of $U(N_c)$ and requiring the full amplitude to vanish generates 
relations between the partial amplitudes known as decoupling identities.      For example letting leg $1$ be a $U(1)$ gluon and examining the 
coefficient of $\tr[T^2T^3\cdots T^n]$ we obtain
\begin{equation}
A_{n:2}^{(2)}(1 ; 2,3,\cdots,n)  +A_{n:1}^{(2)}(1,2,3,\cdots, n ) +A_{n:1}^{(2)}(2,1,3,\cdots, n )+\cdots +A_{n:1}^{(2)}(2,\cdots, 1, n )  =0 \,.
\end{equation}
This allows $A_{n:2}^{(2)}$ to be expressed  in terms of the 
$A_{n:1}^{(2)}$.    Similarly the $A_{n:1,s}^{(2)}$ and $A_{n:1,1}^{(2)}$ may be expressed in terms of the $A_{n:1}^{(2)}$ and $A_{n:r}^{(2)}, r> 2$. 
The decoupling identities can be used iteratively to express the sub-sub leading $SU(N_c)$ terms $A_{n:s,t}^{(2)}$, $s=1,2$, 
in terms of $A_{n:1}^{(2)}$ and $A_{n:r}^{(2)}, r> 2$.  Although this may not be the most efficient expressions for these.    
Finally, if we consider $A_{n:1B}^{(2)}$,
the decoupling identities provide consistency constraints but do not relate these to the other amplitudes:
\begin{equation}
A_{n:1B}^{(2)}(1,2,3,\cdots, n ) +A_{n:1B}^{(2)}(2,1,3,\cdots, n )+\cdots+ A_{n:1B}^{(2)}(2,\cdots, 1, n )=0\,.
\label{eq:decoupleB}
\end{equation}
Decoupling identities do not exhaust the color relations and further constraints arise 
from recursive approaches~\citep{Edison:2011ta,Edison:2012fn}
which imply  extra relations 
involving both $A_{n:1B}^{(2)}$ and other amplitudes. For $n=5$ these contain sufficient information  to determine $A_{5:1B}^{(2)}$ but at $n=6$ 
and beyond the $A_{n:1B}^{(2)}$ is a further function which must be determined.  

In summary, the minimal set of color trace amplitudes which must be determined to fully specify the amplitude are 
$A_{n:1}^{(2)}$, $A_{n:r}^{(2)}$ with $r >2$,  $A_{n:s,t}^{(2)}$ with $s >2$ and $A_{n:1B}^{(2)}$.  

At six-point all partial amplitudes can be expressed in terms of $A_{6:1}^{(2)}$, $A_{6:3}^{(2)}$, $A_{6:4}^{(2)}$ and $A_{6:1B}^{(2)}$.  
Explicitly, the  specifically $U(N_c)$ amplitudes 
are given by
\begin{align}
A_{6:2}^{(2)}( 1;2,3,4,5,6 )&=-A_{6:1}^{(2)}( 1,2,3,4,5,6 )-A_{6:1}^{(2)}( 2,1,3,4,5,6 )-A_{6:1}^{(2)}( 2,3,1,4,5,6 )\notag\\
&-A_{6:1}^{(2)}( 2,3,4,1,5,6 )-A_{6:1}^{(2)}( 2,3,4,5,1,6 )\,,
\notag \\
A^{(2)}_{6:1,1}( 1;2;3,4,5,6 )&=-A^{(2)}_{6:3}( 1,2;3,4,5,6 )+\sum_{\sigma\in OP\{\bar\alpha\}\{\beta\}}A^{(2)}_{6:1}(\sigma)
\notag
\intertext{and}
A^{(2)}_{6:1,2}(1;2,3;4,5,6)&=-A^{(2)}_{6:4}(1,2,3;4,5,6)-A^{(2)}_{6:4}(2,1,3;4,5,6)\notag\\
&-A^{(2)}_{6:3}(2,3;1,4,5,6)-A^{(2)}_{6:3}(2,3;4,1,5,6)-A^{(2)}_{6:3}(2,3;4,5,1,6).
 \label{eq:U1decouple}
 \end{align}
 where $\{\bar\alpha\}=\{2,1\}$, $\{\beta\}=\{3,4,5,6\}$ and $OP\{S_1\}\{S_2\}$ is the set of all mergers of $S_1$ and $S_2$ which preserves the 
 order of $S_1$ and $S_2$ within the merged list. Note the first element in these sums has the list reversed although for a set of two legs this is meaningless. 
The remaining $SU(N_c)$ partial amplitude is given by  
\begin{align}
    &A^{(2)}_{6:2,2}(1,2;3,4;5,6)=-\sum_{Z_2(\mu)}\;\sum_{\eta\;\in OP\{1\}\{\nu\}}
    \;\sum_{\eta'\in OP\{2\}\{\rho\}} A^{(2)}_{6:4}(\eta;\eta')
    \notag\\
    &-\sum_{Z_2(\nu,\rho)}\;\sum_{\sigma\in OP\{\bar\mu\}\{\nu\}}
    A^{(2)}_{6:3}(\rho;\sigma)
    -\sum_{Z_2(\mu)}\sum_{Z_2(\{\nu,\rho\})}\;
    \sum_{\sigma\in OP\{2\}\{\rho\}}A^{(2)}_{6:1,2}(1;\nu;\sigma),
    \label{eq:a622dec}
\end{align}
where $\{\bar\mu\}=\{2,1\}$, $\{\nu\}=\{3,4\}$ and $\{\rho\}=\{5,6\}$.
This is an inefficient expression with considerable cancellation amongst the terms on the RHS. For example, the RHS of the above contains 
terms with double poles in complex momenta whilst $A^{(2)}_{6:2,2}$ does not.

We calculate all eight $U(N_c)$ functions directly and we use (\ref{eq:decoupleB}), (\ref{eq:U1decouple})  and (\ref{eq:a622dec}) as consistency checks.

\section{Structure of the Amplitudes}

The IR singular structure of a color partial amplitude is determined by general theorems~\cite{Catani:1998bh}. Consequently we can split the amplitude into
singular terms $U^{(2)}_{n:\lambda}$ and finite terms $F^{(2)}_{n:\lambda}$, 
\begin{eqnarray}
\label{definitionremainder}
A^{(2)}_{n:\lambda} =& U^{(2)}_{n:\lambda}
+  \; F^{(2)}_{n:\lambda}  +   {\mathcal O}(\epsilon)\, .
\end{eqnarray}
As the all-plus
tree amplitude vanishes, $U^{(2)}_{n:\lambda}$ simplifies considerably and is at worst $1/\epsilon^2$
~\cite{Kunszt:1994np}. 
Specifically, $U^{(2)}_{n:1}$ is proportional to the one-loop amplitude,
\begin{equation}
U_{n:1}^{(2)}  =A_{n:1}^{(1)} \times  
%
\left[ - \sum_{i=1}^{n} \frac{1}{\epsilon^2} \left(\frac{\mu^2}{-s_{i,i+1}}\right)^{\epsilon}  \right] 
\end{equation}
and the two-loop IR divergences for the other un-renormalised partial amplitudes are presented
in a color trace basis in  ref.~\cite{Dunbar:2019fcq}. 

%
The finite remainder function $F_{n:\lambda}^{(2)}$ can be split into polylogarithmic and rational pieces,
\begin{equation}
F_{n:\lambda}^{(2)} = \Fcc_{n:\lambda}+R_{n:\lambda}^{(2)}\; .
\end{equation}
We calculate the former piece using four-dimensional unitarity and the latter using recursion. 
\def\plF{{\rm F}}

The one-loop all-plus amplitude is rational to leading order in $\epsilon$ and in four-dimensional unitarity effectively provides an additional 
on-shell vertex~\citep{Dunbar:2016cxp,Dunbar:2017nfy}.   The two-loop cuts effectively become one-loop cuts with a single insertion of this vertex
which 
yield~\footnote{The functions $\plF^{\rm 2m }$ and $\plF^{\rm 1m }$ are the polylogarithimc parts of two-mass easy and one-mass one-loop box functions respectively.}
\begin{equation}
P_{n:\lambda}^{(2)}   =  \sum_{i}   c^\lambda_{i}  \plF^{2m}_{i}\,,
\end{equation}
where $c^\lambda_i$ are rational coefficients,
\begin{eqnarray}
\plF^{\rm 2m }(S,T,K^2_2,K_4^2)  &=& 
  \Li_2\left(1-{ K_2^2 \over S }\right)
   +\ \Li_2\left(1-{K_2^2 \over T}\right)
 + \Li_2\left(1-{ K_4^2 \over S }\right)
   + \ \Li_2\left(1-{K_4^2 \over T}\right)  
\notag \\ 
& &  -\Li_2\left(1-{K_2^2K_4^2\over S T}\right)
   +{1\over 2} \ln^2\left({ S  \over T}\right)
\end{eqnarray}
and, in the specific case where $K_2^2=0$, 
\begin{eqnarray}
\plF^{\rm 2m }(S,T,0,K_4^2) &\equiv & \plF^{\rm 1m }(S,T,K_4^2)
\notag\\  &=& 
 \Li_2\left(1-{ K_4^2 \over S }\right)
   + \ \Li_2\left(1-{K_4^2 \over T}\right)  
   +{1\over 2} \ln^2\left({ S  \over T}\right)+{\pi^2 \over 6} \; .
\end{eqnarray}
%

\def\Ctmi{C^{2m}_{i}}
\def\Ctmii{C^{2m}_{ii}}
\def\Comoi{C^{1m}_{6:1i}}
\def\Comoii{C^{1m}_{6:1ii}}
\def\Comtii{C^{1m}_{6:2ii}}
\def\Comtiii{C^{1m}_{6:2iii}}
\def\Comthi{C^{1m}_{6:3i}}
\def\Comthii{C^{1m}_{6:3ii}}
\def\Comthiii{C^{1m}_{6:3iii}}
\def\Ctmi{C^{2m}_{a}}
\def\Ctmii{C^{2m}_{b}}
\def\Comoi{C^{1m}_{a}}
\def\Comoii{C^{1m}_{b}}
\def\Comtii{C^{1m}_{c}}
\def\Comtiii{C^{1m}_{d}}
\def\Comthi{C^{1m}_{e}}
\def\Comthii{C^{1m}_{f}}
\def\Comthiii{C^{1m}_{g}}
%

Defining\footnote{Here  a null momentum is represented as a
pair of two component spinors $p^\mu =\sigma^\mu_{\alpha\dot\alpha}
\lambda^{\alpha}\bar\lambda^{\dot\alpha}$. 
We are using a spinor helicity formalism with the usual
spinor products  $\spa{a}.{b}=\epsilon_{\alpha\beta}
\lambda_a^\alpha \lambda_b^{\beta}$  and 
 $\spb{a}.{b}=-\epsilon_{\dot\alpha\dot\beta} \bar\lambda_a^{\dot\alpha} \bar\lambda_b^{\dot\beta}$. 
\noindent{Also}
 $ s_{ab}=(k_a+k_b)^2=\spa{a}.b \spb{b}.a=\la a|b|a]$, $t_{abc}=(k_a+k_b+k_c)^2$, $[a|P_{bc}|d\ra =[ab]\la bd\ra+ [ac]\la cd\ra$ etc.,
$ {\rm tr}_-[ijkl]\equiv {\rm tr}( \frac{(1-\gamma_5)}{2} \ksl_{i} \ksl_{j} \ksl_{k} \ksl_{l} ) =\spa{i}.{j}\spb{j}.{k}\spa{k}.{l}\spb{l}.{i}
\equiv \la i|jkl|i]$ ,\\
\noindent{
${\rm tr}_+[ijkl]\equiv {\rm tr}( \frac{(1+\gamma_5)}{2} \ksl_{i} \ksl_{j} \ksl_{k} \ksl_{l} ) =\spb{i}.{j}\spa{j}.{k}\spb{k}.{l}\spa{l}.{i}$
and $\epsilon(i,j,k,l)={\rm tr}_+[ijkl]-{\rm tr}_-[ijkl]$.}
}
\begin{align}
   \Ctmi(a;b,c;d;e,f) &=
   \frac{i}{3}
   \frac{\spb{e}.f^2}{\spa{a}.b\spa{b}.c\spa{c}.d\spa{d}.a}
   \times \plF^{2m}(t_{abc},t_{bcd},s_{bc},s_{ef})
   \notag\\
   \Ctmii(a;b,c;d;e,f) &=
   \frac{i}{3}
   \frac{\spb{e}.f^2}{\spa{a}.b\spa{b}.d\spa{d}.c\spa{c}.a}
   \times \plF^{2m}(t_{abc},t_{bcd},s_{bc},s_{ef}).
   \label{eq:2mcoefs}
\end{align}
and
\begin{align}
    \Comoi(a,b,c;d,e,f)&=
    \frac{i}{3}
    \frac{t_{abc}\la c|dP_{abc}|a\ra+\la c|defP_{def}|a\ra+\la a|fP_{def}|c\ra s_{ef}}
    {\spa{a}.b\spa{b}.c\spa{c}.a\spa{c}.d\spa{d}.e\spa{e}.f\spa{f}.a}\times \plF^{1m}(s_{ab},s_{bc},t_{def})
    \notag\\
    \notag\\
    \Comoii(a,b,c;d,e,f)&=
     \frac{i}{3}
    \left(\frac{\la a|dP_{abc}|c\ra\la c|dP_{abc}|a\ra+\spa{c}.a(s_{ef}\la a|fP_{abc}|c\ra-\la a|P_{abc}\,efd|c\ra)}
    {\spa{a}.b\spa{b}.c\spa{c}.a\spa{a}.d\spa{d}.c\spa{c}.e\spa{e}.f\spa{f}.a}\right)
    \notag \\
    & \hskip 10.0truecm\times \plF^{1m}(s_{ab},s_{bc},t_{def})
    \notag\\
    \notag\\
    \Comtii(a,b,c;d,e,f)&=
    - i\,
    \frac{\spa{c}.a[d|ef|d]-[d|P_{abc}|c\ra[d|P_{abc}|a\ra}
    {\spa{a}.b\spa{b}.c\spa{c}.a\spa{c}.e\spa{e}.f\spa{f}.a}\times \plF^{1m}(s_{ab},s_{bc},t_{def})
    \notag\\
    \notag\\
    \Comtiii(a,b,c;d,e,f)&=
    i\,
    \frac{[d|P_{abc}|a\ra[d|f|c\ra-[d|P_{abc}|c\ra[d|e|a\ra}
    {\spa{a}.b\spa{b}.c\spa{a}.e\spa{e}.c\spa{c}.f\spa{f}.a}\times \plF^{1m}(s_{ab},s_{bc},t_{def})
    \notag\\
    \notag\\
    \Comthi(a,b,c;d,e,f)&=
    -2i\,
    \frac{t_{abc}^2}
    {\spa{a}.b\spa{b}.c\spa{c}.a\spa{d}.e\spa{e}.f\spa{f}.d}\times \plF^{1m}(s_{ab},s_{bc},t_{def})
    \notag\\
    \notag\\
    \Comthii(a,b,c;d,e,f)&=
    -2i\,
    \frac{[d|P_{abc}|c\ra^2}
    {\spa{a}.b\spa{b}.c\spa{c}.a\spa{c}.e\spa{e}.f\spa{f}.c}\times \plF^{1m}(s_{ab},s_{bc},t_{def})
    \notag\\
    \notag\\
    \Comthiii(a,b,c;d,e,f)&=
    -2i\,
    \frac{\spb	{d}.e^2\spa{c}.a^2}
    {\spa{a}.b\spa{b}.c\spa{c}.a\spa{a}.c\spa{c}.f\spa{f}.a}\times \plF^{1m}(s_{ab},s_{bc},t_{def}).
    \label{eq:1mcoefs}
\end{align}
\def\Nc{N_c}
Note that these six-point coefficients are conformally invariant: a feature noticed for the five-point all-plus amplitude in ref.~\citep{Henn:2019mvc}.

Using these definitions the results for $P_{6:\lambda}^{(2)}$ are: 
\begin{align}
    P_{6:1}^{(2)}(a,b,c,d,e,f)&=\sum_{\mathcal{P}_{6:1}}\Bigl( \Comoi(a,b,c;d,e,f) + \Ctmi(a;b,c;d;e,f)\Bigr),
    \label{eq:poly61}
\\
    P_{6:3}^{(2)}(a,b;c,d,e,f)&
    \notag\\
    =\sum_{\mathcal{P}_{6:3}}\Bigg( &\Comoi(a,b,c;d,e,f)+\Comoi(a,c,b;d,e,f)+\Comoi(c,a,b;d,e,f)
    \notag\\
    &-\Comoii(a,c,d;b,e,f)-\Comoii(c,a,d;b,e,f)-\Comoii(c,d,a;b,e,f)
    \notag\\
    &-\Comoii(d,e,f;c,a,b)+\frac12\Comthiii(d,e,f;a,b,c)
    \notag\\
    &+4\Ctmi(c;d,e;f;a,b)
    +\Ctmi(b;e,f;a;c,d)+\Ctmi(f;b,a;e;c,d)
    \notag\\
    &-\Ctmii(e;f,a;b;c,d)-\Ctmii(f;e,b;a;c,d)
    \notag\\
    &+\Ctmii(d;e,b;f;a,c)-\Ctmi(b;d,e;f;a,c)-\Ctmi(d;e,f;b;a,c)\Bigg),
    \label{eq:poly63}
\end{align}
\begin{align}
    P_{6:4}^{(2)}(a,b,c;d,e,f)&
    \notag\\
    =\sum_{\mathcal{P}_{6:4}}\Bigg(
    &\frac13 \Comthi(a,b,c;d,e,f)-\Comoi(a,b,c;f,e,d)
    \notag\\
    &+\Comoii(d,b,a;c,e,f)+\Comoii(b,d,a;c,e,f)+\Comoii(b,a,d;c,e,f)
    \notag\\
    &+\Ctmi(a;f,e;b;c,d)-\frac{1}{2}\Ctmii(a;b,f;e;c,d)-\frac{1}{2}\Ctmii(f;a,e;b;c,d)
    \notag\\
    &+\Ctmii(a;b,d;c;e,f)-\Ctmi(a;b,c;d;e,f)-\Ctmi(d;a,b;c;e,f)\Bigg),
    \label{eq:poly64}
\end{align}
\begin{align}
    P_{6:2,2}^{(2)}(a,b;c,d;e,f)&
    \notag\\
    =\frac12\sum_{\mathcal{P}_{6:2,2}}\Bigg(
    &\Comthiii(a,b,c;e,f,d)+\Comthiii(b,a,c;e,f,d)+\Comthiii(b,c,a;e,f,d)
    \notag\\
    &+6\Ctmi(d;a,b;c;e,f)-3\Ctmii(a;b,c;d;e,f)-3\Ctmii(b;a,d;c;e,f)\Bigg)
    \label{eq:poly622}
\end{align}
and
\begin{align}
    P_{6:1B}^{(2)}(a,b,c,d,e,f)&
    \notag\\
    =\sum_{\mathcal{P}_{6:1}}\Bigg(
    &\Comthii(a,b,c;f,d,e)-\Comthii(c,b,a;d,e,f,)
    \notag\\
    &+\Comthii(b,f,e;a,c,d)+\Comthii(f,b,e;a,c,d)+\Comthii(f,e,b;a,c,d)
    \notag\\
    &-\Comthii(f,b,c;a,d,e)-\Comthii(b,f,c;a,d,e)-\Comthii(b,c,f;a,d,e)
    \notag\\
    &+6\Ctmii(f;b,e;d;a,c)-6\Ctmi(b;f,e;d;a,c)-6\Ctmi(f;e,d;b;a,c)
    \notag\\
    &+6\Ctmi(a;b,c;d;e,f)+3\Ctmi(f;b,c;e;a,d)+3\Ctmi(c;e,f;b;a,d)
    \notag\\
    &-3\Ctmii(b;c,f;e;a,d)-3\Ctmii(c;e,b;f;a,d)\Bigg).
    \label{eq:poly61B}
\end{align}
This expression for $P_{6:1B}^{(2)}$ matches the $n$-point form of $P_{n:1B}^{(2)}$ given in \cite{Dunbar:2020wdh}. 
The $U(N_c)$ pieces are:
\begin{align}
    P_{6:2}^{(2)}(a;b,c,d,e,f)&
    \notag\\
    =\sum_{\mathcal{P}_{6:2}}
    \Bigg(&\Comoii(b,c,d;a,e,f)+\Comtii(b,c,d;a,e,f)-\Comoi(a,b,c;d,e,f)
    \notag\\
    &-\Comoi(b,a,c;d,e,f)-\Comoi(b,c,a;d,e,f)-2\Ctmi(b;c,d;e;f,a)
    \notag\\
    &+\Ctmii(b;c,a;d;e,f)-\Ctmi(a;b,c;d;e,f)-\Ctmi(b;c,d;a;e,f)\Bigg),
    \label{eq:poly62}
\end{align}
\begin{align}
    P_{6:1,1}^{(2)}(a;b;c,d,e,f)&
    \notag\\
    =\sum_{\mathcal{P}_{6:1,1}}
    \Bigg(
    &\Comtiii(c,d,e;a,b,f)-3\Ctmi(c;d,e;f;a,b)
    \notag\\
    &-\Comtii(b,c,d;a,e,f)-\Comtii(c,b,d;a,e,f)-\Comtii(c,d,b;a,e,f)
    \notag\\
    &+3\Ctmi(b;c,d;e;f,a)+3\Ctmi(c;d,e;b;f,a)-3\Ctmii(c;d,b;e;f,a)\Bigg),
    \label{eq:poly611}
\end{align}
and
\begin{align}
    P_{6:1,2}^{(2)}(a;b,c;d,e,f)&
    \notag\\
    =\sum_{\mathcal{P}_{6:1,2}}
    \Bigg(
    &\frac12(\Comtii(c,b,d;a,e,f)+\Comtii(b,c,d;a,e,f)+\Comtii(b,d,c;a,e,f))
    \notag\\
    &+\frac12(\Comtii(c,b,d;a,f,e)+\Comtii(b,c,d;a,f,e)+\Comtii(b,d,c;a,f,e))
    \notag\\
    &-\frac12(\Comthiii(a,d,e;b,c,f)+\Comthiii(d,a,e;b,c,f)+\Comthiii(d,e,a;b,c,f))
    \notag\\
    &-\frac12\Comthiii(d,e,f;b,c,a)-\Comtii(d,e,f;a,b,c)
    \notag\\
    &-\Comtiii(b,d,e;a,c,f)-\Comtiii(d,b,e;a,c,f)-\Comtiii(d,e,b;a,c,f)
    \notag\\
    &+3\Ctmii(c;b,f;e;a,d)+3\Ctmii(b;e,c;f;a,d)
    \notag\\
    &-3\Ctmi(b;e,f;c;a,d)-3\Ctmi(e;b,c;f;a,d)
    \notag\\
    &+3\Ctmi(c;d,e;f;a,b)+3\Ctmi(d;e,f;c;a,b)-3\Ctmii(d;e,c;f;a,b)
    \notag\\
    &-3\Ctmi(a;d,e;f;b,c)-3\Ctmi(d;e,f;a;b,c)+3\Ctmii(d;e,a;f;b,c)\Bigr)\,.
    \label{eq:poly612}
\end{align}
 
\section{Rational Terms}

As $R_{n:\lambda}^{(2)}$ is a rational function we may calculate it using recursion techniques
by performing a complex shift of its external legs \cite{Britto:2005fq,Risager:2005vk} and analysing the singularities of the resultant complex function $R(z)$.  
This is complicated because the amplitude has double poles in complex momenta. 
The leading poles are determined by the amplitude's factorisation but there are no general theorems that determine the subleading poles.
We use color dressed augmented recursion as reviewed in \cite{Dunbar:2017nfy,Dunbar:2019fcq} to overcome the issue of double poles.
This requires generating certain  doubly off-shell currents 
which we present in appendix~\ref{app_currents}. The specific rational pieces are:

\subsection{$\mathbf{R_{6:1}^{(2)}}$}
\begin{align}
R_{6:1}^{(2)}(a,b,c,d,e,f)&= \frac{i}{9}\sum_{\mathcal{P}_{6:1}}
  \frac{G_{6:1}^1+G_{6:1}^2+G_{6:1}^3+G_{6:1}^4+G_{6:1}^5}{\spa{a}.b\spa{b}.c\spa{c}.d\spa{d}.e\spa{e}.f\spa{f}.a}
\intertext{where}
    G_{6:1}^1(a,b,c,d,e,f)&=\frac{s_{cd}s_{df}\la f|a\,P_{abc}|e\ra}
    {\spa{f}.e\,t_{abc}}
    +\frac{s_{ac}s_{cd}\la a|f\,P_{def}|b\ra}
    {\spa{a}.b\,t_{def}},
    \notag\\
    G_{6:1}^2(a,b,c,d,e,f)&=\frac{\spb{a}.b\spb{e}.f}{\spa{a}.b\spa{e}.f}
    \spa{a}.e^2\spa{b}.f^2
    +\frac12\frac{\spb{f}.a\spb{c}.d}{\spa{f}.a\spa{c}.d}
    \spa{a}.c^2\spa{d}.f^2,
    \notag\\
    G_{6:1}^3(a,b,c,d,e,f)&=\frac{s_{df}\spa{f}.a\spa{c}.d\spb{a}.c\spb{d}.f}{t_{abc}},
    \notag\\
    G_{6:1}^4(a,b,c,d,e,f)&=\frac{\la a|be|f\ra t_{abc}}{\spa{a}.f}
    \intertext{and}
    G_{6:1}^5(a,b,c,d,e,f)&=s_{fa} s_{bc} 
    + s_{ac} s_{be}
    + \frac52 s_{af} s_{cd}  
    -8 [a|bcf|a\ra 
    -8 [a|cde|a\ra 
    -\frac12 [a|cdf|a\ra 
    - \frac{11}2 [b|cef|b\ra
\label{eq:leadingrat}
\end{align}
 This was first calculated in \cite{Dunbar:2016gjb} and later presented in an alternative form \cite{Dunbar:2017nfy}. It was subsequently confirmed
 by Badger et.al.~\cite{Badger:2016ozq}.

\subsection{$\mathbf{R_{6:3}^{(2)}}$}
\begin{align}
R_{6:3}^{(2)}(a,b;c,d,e,f)&=
\sum_{\mathcal{P}_{6:3}}
\Bigg[\frac{i}3\Big(H^1_{6:3}(a,b,c,d,e,f)-H^1_{6:3}(a,b,c,d,f,e)\Big)
\notag\\
&+\frac{i}3
\frac{\Big(G_{6:3}^2(a,b,c,d,e,f)+G_{6:3}^3(a,b,c,d,e,f)+G_{6:3}^4(a,b,c,d,e,f)\Big)}
{\spa{a}.b\spa{b}.c\spa{c}.a\spa{d}.e\spa{e}.f\spa{f}.d}
\notag\\
&+\frac{i}{12}
\frac{G_{6:3}^5(a,b,c,d,e,f)}
{\spa{a}.b\spa{b}.c\spa{c}.d\spa{d}.e\spa{e}.f\spa{f}.a}
\Bigg]
\label{eq:r63full}
\end{align}
where
\begin{align}
H_{6:3}^1(a,b,c,d,e,f)&=
\frac{G_{6:3}^1(a,b,c,d,e,f)}
{\spa{a}.b\spa{b}.c\spa{c}.d^2\spa{d}.e\spa{e}.f\spa{f}.a}
+ \frac{\spb{c}.d}{\spa{c}.d^2}
\frac{\spa{c}.f \spa{d}.b [b|f|d\ra}
{\spa{a}.b \spa{a}.f \spa{b}.f \spa{d}.e \spa{e}.f}
\notag\\
G_{6:3}^1(a,b,c,d,e,f)&= 
s_{ce} \la c|bf|d\ra - s_{cf} \la c|be|d\ra 
\notag\\
G_{6:3}^2(a,b,c,d,e,f)&=\frac{[d|P_{def}b|a]\la d|fP_{def}|a\ra 
+s_{de}[f|cbd|f\ra 
+[b|df|e]\la b|c P_{abc}|e\ra} 
 {t_{def}}
\notag\\
G_{6:3}^3(a,b,c,d,e,f)&=-\frac{s_{df} \la d|fb|c\ra [c|P_{abc}|e\ra}
{\spa{d}.e  t_{def}}
-
\frac{s_{de} \la f|db|c\ra [c|d|e\ra}
{\spa{e}.f t_{def}}
\notag\\
G_{6:3}^4(a,b,c,d,e,f)&=
-s_{bd}s_{de} - [a|bde|a\ra + [b|cde|b\ra 
-[a|bdf|a\ra 
\notag\\
&+[b|cdf|b\ra 
+[b|cef|b\ra 
-[b|def|b\ra
\notag\\
G_{6:3}^5(a,b,c,d,e,f)&=
-4s_{ac}^2 + 2s_{ab}s_{ad} - 2s_{ac}s_{ad}
+2s_{ab}s_{ae} - 2s_{ac}s_{ae} + 2s_{bd}^2 - 2s_{be}^2
+2s_{bf}^2
\notag\\
&- 8s_{ac}s_{cd} + 4s_{bc}s_{cd} + 
 12 s_{bd}s_{cd} + 6 s_{cd}^2 - 8 s_{ac} s_{ce} + 
 12 s_{bc} s_{ce} + 16 s_{bd} s_{ce}
 \notag\\
 &+ 4 s_{be} s_{ce} + 
 8 s_{cd} s_{ce} + 2 s_{c e}^2 + 2 s_{c f}^2 - 8 s_{a c} s_{d e} - 
 4 s_{a d} s_{d e} - 4 s_{b c} s_{d e} + 4 s_{c d} s_{d e}
 \notag\\
 & +4 s_{c e} s_{d e} - 8 [a|bce|a\ra 
 - 39 [a|bcf|a\ra - 
 18 [a|bdf|a\ra + 2 [a|bef|a\ra 
 \notag\\
 &- 10 [a|cdf|a\ra  - 2 [a|cef|a\ra - 4 [a|def|a\ra + 8 [b|cde|b\ra - 
 4 [b|cdf|b\ra 
 \notag\\
 &- 4 [b|cef|b\ra - 4 [b|def|b\ra - 
 4 [c|def|c\ra
 \label{eq:r63pieces}
\end{align}
\subsection{$\mathbf{R_{6:4}^{(2)}   }$}
\begin{align}
 R_{6:4}^{(2)}(a,b,c,d,e,f)&=\frac{i}{36}\sum_{\mathcal{P}_{6:4}}
    \Bigg[\frac{\Big(G_{6:4}^{1}(a,b,c,d,e,f)+G_{6:4}^{2}(a,b,c,d,e,f)\Big)
 }
 {\spa{a}.b\spa{b}.c\spa{c}.a\spa{d}.e\spa{e}.f\spa{f}.d}
 \notag\\
 &\hspace{1.3cm}+12\frac{\Big(G^{3}_{6:4}(a,b,c,d,e,f)+G^{4}_{6:4}(a,b,c,d,e,f)\Big)}{\spa{a}.b\spa{c}.d\spa{d}.e\spa{e}.f\spa{f}.c}
 \Bigg],
\end{align}
 where
 \begin{align}
 G^{1}_{6:4}(a,b,c,d,e,f)&=\frac{4\,\la e|P_{abc}a|b\ra[e|dP_{abc}|b]}{t_{abc}},
 \notag\\
 G^{2}_{6:4}(a,b,c,d,e,f)&= s_{ad}^2 + 106\,s_{ab}s_{ad} + 102\,[a|bcd|a\ra - 4\,[a|bde|a\ra - 4\,[a|dbe|a\ra,
 \notag\\
 G^{3}_{6:4}(a,b,c,d,e,f)&=-\frac{\spb{a}.b}{\spa{a}.b}\Bigl(\la a|cd|b\ra+\la a|ef|b\ra\Bigr),
 \notag\\
 G^{4}_{6:4}(a,b,c,d,e,f)&=[a|cd|b]+[a|ef|b].
\end{align}
\subsection{$\mathbf{R_{6:2,2}^{(2)}}$}
\begin{align}
R_{6:2,2}^{(2)}(a,b;c,d;e,f)&=\sum_{\mathcal{P}_{6:2,2}}
    i\,\frac{G_{6:2,2}^{1}(a,b,c,d,e,f)+G_{6:2,2}^{2}(a,b,c,d,e,f)
 }
 {\spa{a}.b\spa{b}.c\spa{c}.a\spa{d}.e\spa{e}.f\spa{f}.d},
\end{align}
where
\begin{align}
G_{6:2,2}^{1}(a,b,c,d,e,f)&= \frac{\la b|P_{abc}f|d\ra[b|cP_{abc}|d] } 
 {t_{abc}},
 \notag\\
 G_{6:2,2}^{2}(a,b,c,d,e,f)&=s_{ad}[e|P_{bc}|e\ra -s_{ac}[e|P_{fa}|e\ra -s_{af}s_{ae} -s_{ae}s_{cd}.
\end{align}

\subsection{$\mathbf{R_{6:1B}^{(2)}}$}
An $n$-point formula was conjectured in \cite{Dunbar:2020wdh} and we find agreement.
\begin{equation}
R_{6:1B}^{(2)}(a,b,c,d,e,f)=R_{6:1B_1}^{(2)}(a,b,c,d,e,f)+R_{6:1B_2}^{(2)}(a,b,c,d,e,f)
\end{equation}
where
\begin{equation}
 R_{6:1B_1}^{(2)}(a,b,c,d,e,f) =
 {-2i \over {\CY(a, b, c, d, e, f) } }
 \times\sum_{a\leq i < j < k < l \leq f } \epsilon(i,j,k,l)
\end{equation}
and
\begin{eqnarray}
  & &R_{6:1B_2}^{(2)}(a,b,c,d,e,f)  =4i
 \Bigl(
{  \epsilon(c, d, e, f) \over \CY(a, b, d, e, c, f)  } + 
{   \epsilon(c, d, e, f)\over  \CY(a, b, e, c, d, f) }+ 
 {  \epsilon(c, d, e, f)\over  \CY(a, b, e, d, c, f)  }  
 \notag\\
 & &+ { \epsilon(a, b, c, d)\over  \CY(a, c, d, b, e, f)  } 
 -{ \epsilon(a, b, c, f)\over  \CY(a, c, d, e, b, f)  } + 
  { \epsilon(a, b, c, d)\over  \CY(a, d, b, c, e, f)  }  
 -{ \epsilon(a, c, d, f)\over \CY(a, d, b, e, c, f)  }  
 \notag \\
 & &  +{ \epsilon(a, b, c, d) \over \CY(a, d, c, b, e, f)  } + 
  { \epsilon(a, b, d, f)\over \CY(a, d, c, e, b, f)  }  
 -{ \epsilon(a, c, d, f)\over \CY(a, d, e, b, c, f)  } + 
  { \epsilon(a, b, d, f)\over \CY(a, d, e, c, b, f)  }  
\notag\\
 & & -{ \epsilon(a, d, e, f)\over \CY(a, e, b, c, d, f)  } + 
  { \epsilon(a, c, e, f)\over \CY(a, e, b, d, c, f)  } + 
  { \epsilon(a, c, e, f)\over \CY(a, e, d, b, c, f)  } 
 -{ \epsilon(a, b, e, f)\over \CY(a, e, d, c, b, f) }
\Bigr)
\; , 
\end{eqnarray}
where $\CY$ is the Parke-Taylor denominator,
\begin{equation}
\CY(a, b, c, d, e, f)  = \spa{a}.b \spa{b}.c \spa{c}.d \spa{d}.e \spa{e}.f \spa{f}.a \,.
\end{equation}

\subsection{$\mathbf{R_{6:1,1}^{(2)}}$}
We also calculate the $U(N_c)$ amplitudes
\begin{align}
    R_{6:1,1}^{(2)}(a;b;c,d,e,f) = \sum_{\mathcal{P}_{6:1,1}}&\Bigg(
    i\,\frac{G_{6:1,1}^1(a,b,c,d,e,f)+G_{6:1,1}^2(a,b,c,d,e,f)}
    {\spa{b}.c\spa{c}.d\spa{d}.b\spa{a}.e\spa{e}.f\spa{f}.a}
    \notag\\
    &+i\,\frac{G_{6:1,1}^3(a,b,c,d,e,f)}
    {\spa{a}.c\spa{c}.d\spa{d}.b\spa{b}.e\spa{e}.f\spa{f}.a}\Bigg)
\end{align}
where
\begin{align}
    G_{6:1,1}^1(a,b,c,d,e,f)&=\frac{[c|P_{bcd}\,efP_{bcd}\,b|c\ra}{t_{bcd}},
    \notag\\
    G_{6:1,1}^2(a,b,c,d,e,f)&=2s_{ab} s_{cd}
    -s_{ac}s_{ae}
    +s_{ac}s_{cd}
    +s_{ad} s_{cd}
    -s_{cd}^2
    -s_{cd} s_{ce}
    -s_{cd} s_{cf} 
    -s_{cd} s_{df} 
    \notag\\
    &-[a|cde|a\ra 
    + \frac12 [c|def|c\ra
    \notag 
    \intertext{and}
    G_{6:1,1}^3(a,b,c,d,e,f)&=2 s_{ab} s_{ac}
    +2 s_{ac}^2
    +2 s_{ac} s_{ad}
    +2 s_{ac} s_{ae}
    +2 s_{ac} s_{bc}
    - s_{ae} s_{bc}
    + s_{ab} s_{cd}
    + s_{ac} s_{cd}
    \notag\\
    &+ s_{ad} s_{cd}
    -2 s_{ae} s_{cd} 
    +2 s_{ad} s_{ce} 
    -2 s_{ae} s_{ce}
    - s_{cd} s_{ce}
    - s_{ce}^2
    - s_{cd} s_{cf}
    + s_{ce} s_{df}
    \notag\\
    &-\frac12 s_{cd} s_{ef}
    + 2 [a|cbd|a\ra 
    + 2 [a|cbe|a\ra 
    + 4 [a|cde|a\ra 
    - [c|def|c\ra.
\end{align}

\subsection{$\mathbf{R_{6:1,2}^{(2)}}$}
\begin{align}
    R_{6:1,2}^{(2)}(a;b,c;d,e,f) &= \sum_{\mathcal{P}_{6:1,2}}
    \frac{i\Big(G_{6:1,2}^1(a,b,c,d,e,f)+G_{6:1,2}^2(a,b,c,d,e,f)\Big)}
    {\spa{e}.f\spa{f}.a\spa{a}.e\spa{b}.c\spa{c}.d\spa{d}.b}
\end{align}
where
\begin{align}
   G_{6:1,2}^1(a,b,c,d,e,f)&= -\frac{[e|fP_{bcd}dbP_{bcd}|e\ra+[e|P_{bcd}bcP_{bcd}a|e\ra}{t_{bcd}},
   \notag\\
   G_{6:1,2}^2(a,b,c,d,e,f)&=[a|bce|a\ra-2[b|dce|b\ra+[b|def|b\ra \,.
\end{align}

\subsection{$\mathbf{R_{6:2}^{(2)}}$}
$\mathbf{R_{6:2}^{(2)}}$ is compactly written by its decoupling identity which we have checked numerically:

\begin{align}
   R_{6:2}^{(2)}(a;b,c,d,e,f)=&-R_{6:1}^{(2)}(a,b,c,d,e,f)-R_{6:1}^{(2)}(b,a,c,d,e,f)-R_{6:1}^{(2)}(b,c,a,d,e,f)
   \notag \\ &
   -R_{6:1}^{(2)}(b,c,d,a,e,f)-R_{6:1}^{(2)}(b,c,d,e,a,f)\,.
\end{align}

These expressions are valid for both $U(N_c)$ and $SU(N_c)$ gauge groups and are remarkably compact.
We have confirmed that they satisfy the constraints arising from the decoupling identities.
The $SU(N_c)$ amplitudes have the correct collinear limits: 
all non-adjacent and inter-trace 
limits vanish and adjacent limits within a single trace factorize correctly.
All of the partial amplitudes have the correct symmetries.
Recursion involves choosing specific legs to shift, breaking the symmetry of the amplitude. Restoration of 
this symmetry is a powerful check of the validity of our results. 
We have checked that none of the $R_{6:\lambda}^{(2)}$ are annihilated by the conformal operator. 

\vfill\eject

\section{Conclusions}
Computing perturbative gauge theory amplitudes to high orders is an important but difficult task. 
In this article, we have calculated the full color all-plus six-point two-loop amplitude and presented the results in simple analytic forms.  
We have computed all the color components directly thus presenting the first complete six gluon two-loop scattering amplitude. 

Our methodology obtains these results bypassing the need to determine two-loop non-planar integrals.  
There are some inherent assumptions in our methods however,  the results satisfy a variety of consistency checks. Firstly, they give the correct results for the five-point amplitudes and for $A_{6:1}^{(2)}$ which was computed subsequently.  Secondly, we have generated the full set of amplitudes and then checked  the decoupling identities are satisfied.  We have checked the collinear limits of the amplitudes. Note that 
the singular terms $U_{n:\lambda}^{(2)}$ and the polylogarithms $P_{n:\lambda}^{(2)}$ must combine to give the correct collinear limits as in
ref~\cite{Dunbar:2016aux,Dunbar:2016cxp}.   

Analytic forms are particularly useful in studying formal properties of amplitudes.
For example we have confirmed that the coefficients of the polylogarithms are conformally invariant whilst the rational terms are not.

\def\COMMENT{
\begin{figure}[h]
\begin{center}
\includegraphics[scale=0.8]{fig6.eps}
\end{center}
\caption{Diagram showing the tree-(two-loop) factorisation. Color dressing this diagram has no Tr$(a S_1)$Tr$(b S_2)$ contributions.}
\label{fig:axialg}
\end{figure} 
}%

\section{Acknowledgements}
DCD was supported by STFC grant ST/L000369/1. JMWS was 
supported by STFC grant ST/S505778/1. ARD was 
supported by an STFC studentship.

\appendix

\section{Currents and Recursion}
\label{app_currents}

Augmented recursion was reviewed in \cite{Dunbar:2019fcq} and shown to work for a full color amplitude. We will outline the steps here.
The amplitude contains double poles and so factorisation theorems don't provide the full pole structure. Mathematically we can take the residue of a function via its Laurent expansion
\begin{eqnarray}
f(z) &=& \frac{c_{-2}}{(z-z_{j})^2}+ \frac{c_{-1}}{(z-z_{j})}+\mathcal{O}((z-z_{j})^0)\; , \notag \\
\end{eqnarray}
where the residue is simply
\begin{equation}
{\rm Res}\Bigl[ {f(z)\over z}\Bigr]\Bigr\vert_{z_j}  = 
-\;\frac{c_{-2}}{z_{j}^2}+ \frac{c_{-1}}{z_{j}}\,.
\end{equation}
As $R_{n:\lambda}^{(2)}$ is a rational function we can obtain it recursively
by performing a complex shift of its external legs \cite{Britto:2005fq,Risager:2005vk} and analysing the singularities of the resultant complex function $R(z)$.

Here $z$ is a complex parameter introduced by the shift and the shift must be chosen carefully so that $R(z)$ vanishes for large $|z|$. Cauchy's theorem then tells us 
\begin{equation}
R= R(0)= -\sum_{z_j\neq 0} {\rm Res}\Bigl[ {R(z)\over z}\Bigr]\Bigr\vert_{z_j}\;.
 \label{AmpAsResidues2}
\end{equation}
For tree amplitudes this can be achieved by the Britto-Cachazo-Feng-Witten shift \cite{Britto:2005fq}. 
For the two-loop all-plus amplitude the Risager shift \cite{Risager:2005vk}
\begin{equation}
    \begin{split}
         \lambda_a\to\lambda_{\hat{a}}=\lambda_a+z\spb{b}.c\lambda_\eta\,,\\
         \lambda_b\to\lambda_{\hat{b}}=\lambda_b+z\spb{c}.a\lambda_\eta\,,\\
         \lambda_c\to\lambda_{\hat{c}}=\lambda_c+z\spb{a}.b\lambda_\eta\,,
   \end{split}
   \label{Kasper}
\end{equation}
preserves overall momentum conservation and gives the desired large $|z|$ behaviour, where $\lambda_\eta$ must satisfy $\spa{a}.{\eta}\neq 0$ etc. but is otherwise unconstrained. Shifting the legs breaks the symmetry of the amplitude so recovering the necessary symmetries (the cyclic symmetries as well as $\lambda_\eta$ independence) provides a strong check. The symmetry is recovered by the Risager shift.

The leading poles are determined by the amplitude's factorisation but there are no general theorems that determine the subleading poles.
The Risager shift excites poles corresponding to tree:two-loop  and one-loop:one-loop factorisations. The former involve only single poles and their contributions
are readily obtained from the rational parts of the five-point two-loop amplitude~\cite{Badger:2019djh,Dunbar:2019fcq}:

\begin{align}
    R^{(2)}_{5:1}(a^+,b^+,c^+,d^+,e^+)&=\frac{i}{9}\frac1{\spa{a}.b\spa{b}.c\spa{c}.d\spa{d}.e\spa{e}.a}
    \sum_{S_{5:1}}\Bigl(\frac{{\rm tr}^2_+[deab]}{s_{de}s_{ab}}+5s_{ab}s_{bc}+s_{ab}s_{cd}\Bigr)\,,
\notag\\
    R^{(2)}_{5:3}(a^+,b^+;c^+,d^+,e^+)&=-\frac{2i}3\frac1{\spa{a}.b\spa{b}.a\spa{c}.d\spa{d}.e\spa{e}.c}
    \sum_{S_{5:3}}\left(\frac{{\rm tr}_-[acde]{\rm tr}_-[ecba]}{s_{ae}s_{cd}}+\frac32s_{ab}^2\right)
    \notag
\intertext{and}
    R^{(2)}_{5:1B}(a^+,b^+,c^+,d^+,e^+)&=2i\epsilon\left(a,b,c,d\right)\sum_{Z_5(a,b,c,d,e)}\text{C}_{\text{PT}}(a,b,e,c,d).
\end{align}

\begin{figure}[h]
\begin{center}
\includegraphics[scale=0.8]{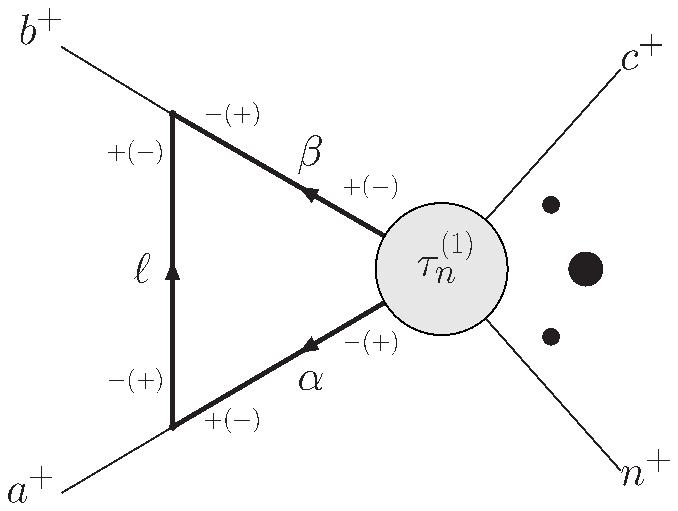}
\end{center}
\caption{Diagram containing the leading and sub-leading poles as $s_{ab}\to 0$. The axial gauge construction permits the off-shell continuation of the internal 
legs.}
\label{fig:axialg}
\end{figure}

The one-loop:one-loop factorisations involve double poles and we need to determine the sub-leading pieces.
By considering a diagram of the form fig.~\ref{fig:axialg} using an axial gauge formalism \cite{Kosower:1989xy,Schwinn:2005pi}, we can determine the full pole structure of the rational piece, 
including the non-factorising simple poles.
We have used this approach previously to compute one-loop~\cite{Dunbar:2010xk,Alston:2015gea,Dunbar:2016dgg}
and two-loop amplitudes~\cite{Dunbar:2016aux,Dunbar:2016gjb,Dunbar:2017nfy,Dunbar:2019fcq},
we labelled this process  {\em augmented recursion}. 

The principal helicity assignment in fig.~\ref{fig:axialg}, gives
\begin{eqnarray}
\int\!\! d {\Lambda}^{c}(\alpha^+,a^+,b^+,\beta^-)\;\tau_{n}^{(1),c} (\alpha^{-},\beta^{+},c^+,...,n^+)
\label{eq:tauintinitA}
\end{eqnarray}
where 
\begin{equation}
\int\!\! d\Lambda^c(\alpha^+,a^+,b^+,\beta^-)  \equiv \frac{i}{c_\Gamma(2\pi)^D}\int\!\! \frac{d^D\ell}{\ell^2\alpha^2\beta^2} 
{\cal V}_3(\alpha,a,\ell) {\cal V}_3 ( \ell,b,\beta) \,,
\end{equation}
the vertices are in axial gauge and $\tau_{n}^{(1),c}$ is a doubly off-shell current where $c$ denotes an implicit sum over color.

As we are only interested in the residue on the $s_{ab}\to 0$ pole, we do not need the exact current.
It is sufficient that the approximate current satisfies two conditions~\cite{Alston:2015gea,Dunbar:2016aux}:
\begin{enumerate}
    \item[(C1)] The current contains the leading singularity as $s_{\alpha\beta}\to0$ with $\alpha^2,\beta^2\neq0$,
    \item[(C2)] The current is the one-loop, single-minus amplitude in the on-shell limit $\alpha^2,\beta^2\to0$, $s_{\alpha\beta}\neq0$.
\end{enumerate} 
This process is detailed in~\cite{Dunbar:2017nfy} and applied to the full color case in~\cite{Dunbar:2019fcq}.

The $U(N_c)$ color decomposition of $d\Lambda^c$ contains a common kinematic factor so we have the color decompositions
\begin{equation}
\tau_n^{(1),c}  = \sum _{\lambda} C_{\lambda} \tau^{(1)}_{n:\lambda}  
\;\;\; {\rm and} \;\; 
\int d \Lambda^c =  C_{\Lambda} \int d \Lambda_0 
\end{equation}
where 
\begin{equation}
\int d \Lambda_0(\alpha^+,a^+,b^+,\beta^-) =\frac{i}{(2\pi)^D}\int\!\! \frac{d^D\ell}{\ell^2\alpha^2\beta^2} 
\frac{[a|\ell|q\ra[b|\ell|q\ra }{\spa a.q\spa b.q} \frac{\spa \beta.q^2}{\spa\alpha.q^2}\,.
\end{equation}
Hence the full color contribution is
\begin{eqnarray}
\sum_{\lambda} C_{\Lambda} C_{\lambda} \int\!\! d \Lambda_0 (\alpha^+,a^+,b^+,\beta^-)\;
\tau_{n:\lambda}^{(1)} (\alpha^{-},\beta^{+},c^+,\cdots n^+).
\label{eq:fullcolortau}
\end{eqnarray}

The various $\tau^{(1)}_{n:\lambda}$ 
can be expressed as  sums of the leading amplitudes $\tau^{(1)}_{n:1}$ via a series of $U(1)$ decoupling identities.
For the six-point case there are three currents to calculate.
 $\tau^{(1)}_{6:1}(\alpha^-,\beta^+,c^+,d^+,e^+,f^+)$ has been calculated 
previously~\cite{Dunbar:2016gjb} and presented for arbitrary $q$ \cite{Dunbar:2017nfy}.  
The remaining two currents are given by 
\begin{align}
       &\tau^{(1)}_{6:1}(\alpha^-,c^+,\beta^+,d^+,e^+,f^+) 
       \notag\\
       &=\frac{i}3
       \Bigg(
       \frac{\spa{d}.f \spa{\alpha}.e^3 \spb{e}.f}
       {\spa{c}.{\beta} \spa{d}.e^2 \spa{e}.f^2 \spa{\alpha}.c \spa{\beta}.d}
       - \frac{\spa{\alpha}.d^3 \spa{\beta}.e [d|c|\alpha\ra}
       {\spa{c}.{\beta} \spa{d}.e^2 \spa{e}.f \spa{f}.{\alpha} \spa{\alpha}.c \spa{\beta}.d^2}
       \notag\\
       &+\frac{[c|d|\alpha\ra^3}
       {\spa{e}.f \spa{f}.{\alpha} \spa{\beta}.d^2 [c|P_{\beta d}|e\ra t_{c \beta d}}
       +\frac{[f|c|\alpha\ra^3}
       {\spa{c}.{\beta} \spa{d}.e^2 \spa{\alpha}.c [f|c|\beta\ra [c|P_{\alpha\beta}|c\ra} 
       \notag\\
       &+\frac{\spb{c}.f^3}
       {\spb{f}.{\alpha} \spb{\alpha}.c t_{\beta de}}
       \times
       \Bigg[\frac{\spb{\beta}.e}
       {\spa{d}.e \spa{\beta}.d}
       + \frac{\spb{c}.{\beta} \spb{\beta}.d}
       {\spa{d}.e [c|P_{\beta d}|e\ra} 
       -\frac{\spb{d}.e \spb{e}.f}
       {\spa{\beta}.d [f|c|\beta\ra}
      \Bigg]\Bigg) + \mathcal{O}(s_{\alpha\beta})
    \label{eq:nonadj1}
\end{align}
and
\begin{align}
   &\tau^{(1)}_{6:1}(\alpha^-,c^+,d^+,\beta^+,e^+,f^+)
   \notag\\
   &=\frac{i}3
   \Bigg(-\frac{\spa{c}.{\beta} \spa{\alpha}.d^3 \spb{c}.d}
   {\spa{c}.d^2 \spa{d}.{\beta}^2 \spa{e}.f \spa{f}.{\alpha} \spa{\beta}.e}
   +\frac{\spa{\alpha}.e^3 \spa{\beta}.f \spb{e}.f}
   {\spa{c}.d \spa{d}.{\beta} \spa{e}.f^2 \spa{\alpha}.c \spa{\beta}.e^2}
   \notag\\
   &+ \frac{[c|d|\alpha\ra^3}
   {\spa{d}.{\beta}^2 \spa{e}.f \spa{f}.{\alpha} [c|P_{d\beta}|e\ra t_{cd\beta}}
   +\frac{[f|P_{cd}|\alpha\ra^3}
   {\spa{c}.d \spa{\alpha}.c \spa{\beta}.e^2 [f|P_{\alpha c}|d\ra t_{\alpha cd}}
   \notag\\
   &+\frac{\spb{c}.f^3}
   {\spb{f}.{\alpha} \spb{\alpha}.c t_{d\beta e}}
   \times\Bigg[
   \frac{\spb{d}.e}
   {\spa{d}.{\beta} \spa{\beta}.e}
   + \frac{\spb{c}.d \spb{d}.{\beta}}
   {\spa{\beta}.e [c|P_{d\beta}|e\ra}
   -\frac{\spb{e}.f \spb{\beta}.e}
   {\spa{d}.{\beta} [f|P_{\alpha c}|d\ra}
      \Bigg]\Bigg) + \mathcal{O}(s_{\alpha\beta}).
   \label{eq:nonadj2}
\end{align} 
Many of the terms in the non-adjacent currents don't give rationals upon integration. We are thus left with
\begin{align}
    &\int\!\!\frac{d^D\ell}{\ell^2\alpha^2\beta^2} 
       \frac{i}{(2\pi)^D} \frac{[a|\ell|q\ra[b|\ell|q\ra}
       {\spa{a}.q\spa{b}.q}
       \tau^{(1)}_{6:1}(\alpha^-,c^+,\beta^+,d^+,e^+,f^+)|_{\mathbb{Q}}
       \notag\\
       &=\frac{i}6
       \frac{\spb{a}.b} {\spa{a}.b}
       \Bigg(
       \frac{\spa{d}.f \spa{a}.e^3 \spb{e}.f \spa{b}.q^2}
       {\spa{c}.b \spa{d}.e^2 \spa{e}.f^2 \spa{a}.c \spa{b}.d \spa{a}.q^2}
       -\frac{\spa{a}.d^3 \spa{b}.e [d|c|a\ra \spa{b}.q^2}
       {\spa{c}.b \spa{d}.e^2 \spa{e}.f \spa{f}.a \spa{a}.c \spa{b}.d^2 \spa{a}.q^2}
       \notag\\
       &+\frac{[f|c|a\ra^3 \spa{b}.q^2}
       {\spa{c}.b \spa{d}.e^2 \spa{a}.c \spa{a}.q^2 [f|c|b\ra [c|P_{ab}|c\ra}\Bigg)
       \label{eq:nonadjcurrent1}
\end{align}
and
\begin{align}
    &\int\!\!\frac{d^D\ell}{\ell^2\alpha^2\beta^2} 
       \frac{i}{(2\pi)^D} \frac{[a|\ell|q\ra[b|\ell|q\ra}
       {\spa{a}.q\spa{b}.q}
       \tau^{(1)}_{6:1}(\alpha^-,c^+,d^+,\beta^+,e^+,f^+)|_{\mathbb{Q}}
       \notag\\
       &=\frac{i}6
       \frac{\spb{a}.b} {\spa{a}.b}
       \Bigg(
       \frac{\spa{a}.e^3 \spa{b}.f \spa{b}.q^2 \spb{e}.f}
       {\spa{c}.d \spa{d}.b \spa{e}.f^2 \spa{a}.c \spa{a}.q^2 \spa{b}.e^2}
       -\frac{\spa{c}.b \spa{b}.q^2 \spa{a}.d^3 \spb{c}.d}
       {\spa{c}.d^2 \spa{d}.b^2 \spa{e}.f \spa{f}.a \spa{a}.q^2 \spa{b}.e}
\Bigg).
       \label{eq:nonadjcurrent2}
\end{align}
We then color-dress  fig.~\ref{fig:axialg}, sum over all distinct diagrams, extract the contribution to each color structure and take the residues. Summing over all the channels excited by the Risager shift and all helicities gives the full color two-loop amplitude. 
%



\bibliography{TwoLoop}{}

\bibliographystyle{ieeetr}

\end{document}